\documentclass[%
 reprint,
superscriptaddress,
 amsmath,amssymb,
 aps,
]{revtex4-2}

\usepackage[dvipdfmx]{graphicx}
\usepackage{bmpsize}
\usepackage{dcolumn}
\usepackage{bm}
\usepackage[version=4]{mhchem} 
\usepackage{siunitx}
\usepackage{hyperref}

\newcommand{\REV}[1]{{#1}}

\begin{document}

\preprint{APS/123-QED}

\title{Efficient dataset generation for machine learning perovskite alloys}

\author{Henrietta Homm}
\affiliation{%
 Department of Applied Physics, Aalto University, P.O. Box 11100, 00076 Aalto, Finland
}%

\author{Jarno Laakso}
\affiliation{%
 Department of Applied Physics, Aalto University, P.O. Box 11100, 00076 Aalto, Finland
}%

\author{Patrick Rinke}
\email{patrick.rinke@tum.de}
\affiliation{%
 Department of Applied Physics, Aalto University, P.O. Box 11100, 00076 Aalto, Finland
}%
\affiliation{Physics Department, Technical University of Munich, Garching, Germany}
\affiliation{Atomistic Modelling Center, Munich Data Science Institute, Technical University of Munich, Garching, Germany}
\affiliation{Munich Center for Machine Learning (MCML)}

\date{\today}

\begin{abstract}
\REV{Lead-based} perovskite solar cells have reached high efficiencies, but toxicity and lack of stability hinder their wide-scale adoption. These issues have been partially addressed through compositional engineering of perovskite materials, but the vast complexity of the perovskite materials space poses a significant obstacle to exploration. We previously demonstrated how machine learning (ML) can accelerate property predictions for the \ce{CsPb(Cl/Br)3} perovskite alloy. However, the substantial computational demand of density functional theory (DFT) calculations required for model training prevents applications to more complex materials. Here, we introduce a data-efficient scheme to facilitate model training, validated initially on \ce{CsPb(Cl/Br)3} data and extended to the ternary alloy \ce{CsSn(Cl/Br/I)3}. Our approach employs clustering to construct a compact yet diverse initial dataset of atomic structures. We then apply a two-stage active learning approach to first improve the reliability of the ML-based structure relaxations and then refine accuracy near equilibrium structures. Tests for \ce{CsPb(Cl/Br)3} demonstrate that our scheme reduces the number of required DFT calculations during the different parts of our proposed model training method by up to 20\% and 50\%. The fitted model for \ce{CsSn(Cl/Br/I)3} is robust and highly accurate, evidenced by the convergence of all ML-based structure relaxations in our tests and an average relaxation error of only 0.5 meV/atom. 
\end{abstract}

\maketitle


\section{Introduction}

Halide perovskite (\ce{ABX_3} with X = Cl, Br or I) materials have shown great promise in optoelectronic applications. For example, perovskite solar cells (PSCs) have achieved a record power-conversion efficiency of over 26\% \cite{liu2023bimolecularly, liang2023homogenizing, nrelchart} and now almost equal the market-dominating conventional crystalline silicon devices \cite{ma2023developments}. Also perovskite-based light-emitting diodes (PeLEDs) have advanced and now achieve high brightness, high external quantum efficiency, and excellent monochromaticity \cite{lu2019metal, fakharuddin2022perovskite, liu2021metal}. The difficulties hindering the commercialization of halide perovskite materials lie in their instability against external stresses, such as heat, moisture, and oxygen \cite{zhou2019chemical, park2019intrinsic, chen2022critical, hu2021defect}, as well as the toxicity of Pb as the most common B-site element \cite{giustino2016toward, ke2019prospects, konstantakou2017critical, zhang2021lead}.

The flexibility of the perovskite structure for elemental substitutions facilitates property tuning and materials design by compositional engineering \cite{saliba2019polyelemental, lu2019metal, zhang2023composition}. For example, most state of the art PSCs employ cation mixing at the A-site to enhance power conversion efficiency and stability \cite{sun2021data, liang2023homogenizing}. Similarly, the B-site atom can be substituted by different metal ions to adjust various properties of the material \cite{zhang2023composition}. Halide alloying enables further adjustments of light emission wavelength in PeLEDs \cite{ji2021halide, karlsson2021mixed} and optimization of optoelectronic properties in PSCs \cite{xu2023challenges, zhang2023composition}.

Much of PSC research has focused on lead-based materials, which pose environmental concerns due to their toxicity. The substitution of lead with tin represents a promising avenue towards the development of lead-free perovskite photovoltaics \cite{lyu2017addressing, ke2019prospects}, but investigations into tin-based perovskite alloys have been limited. For example, Li et al. employed density functional theory (DFT) to investigate the structural, electronic, and optical properties of \ce{CsSn(Cl/Br/I)3} \cite{li2023first}. Their analysis was, however, limited to specific alloy concentrations realizable in a  5-atom unit cell. In another study, the cluster expansion was employed for a more comprehensive examination of the binary alloys \ce{CsSn(Cl/Br)3}, \ce{CsSn(Cl/I)3}, and \ce{CsSn(Br/I)3} \cite{bechtel2018first}. Notably, the simultaneous mixing of all three halides was not considered, leaving most of the alloy space unexplored.

First-principles calculations, such as DFT, play an important role in compositional engineering strategies, as they provide predictions of the material properties with exact control over material composition and structure. However, the vastness of configurational space and the computational demand of first principles calculations renders a systematic screening of promising materials candidates intractable with DFT alone.

Machine learning (ML) offers an alternative. ML methods are now widely applied in materials science \cite{Himanen/Geurts/Foster/Rinke:2019,Schmidt/Marques/Botti:2019,Chong2023} and have been particularly successful at  predicting the properties of atomic structures quickly and efficiently \cite{ye2018deep, kovacs2023evaluation, stanev2018machine, batzner2022e3, schutt2018schnet}. A good ML model needs high-quality data, as the predictions can only be as accurate as the training data. In materials research, such databases are still often produced with computational methods \cite{jain2013commentary, curtarolo2012aflow} and not experiment due to the comparative ease of generating the necessary data volumes and data standards. Since even computational methods are not infinitely scalable, data generation with, e.g. DFT, is often the bottleneck in the whole ML workflow, although neural network training resources can also become considerable. A pertinent question in this context is how to build training datasets with as few DFT calculations as possible. Decreasing the amount of training data would not only reduce the required computational budget for data generation and model training, but might also reduce model complexity and thus accelerate predictions.

Previously, we generated a dataset of atomic structures, relaxation trajectories and corresponding energies for the  \ce{CsPb(Cl/Br)3} perovskite alloy \cite{laakso2022compositional}. Single point structures in the dataset were generated by randomly varying the perovskite structure and Cl/Br concentration. For a subset of the structures, we performed DFT relaxations and included the structural snapshots from the relaxation trajectories in the dataset. This approach worked well for covering the structural space of a binary alloy, but for more complex materials the required amount of structures, and hence DFT calculations, quickly grows too large. For example, when going from a binary to a ternary alloy in a $2\times2\times2$ supercell including 24 halide atoms, the number of possible compositions already increases from 25 to 325.

\emph{Coreset selection} is a general data reduction strategy (also referred to as dataset pruning), that selectively removes training data while preserving sufficient prediction accuracy \cite{yang2023dataset}. Coreset selection has been successfully implemented in different applications, such as large language models \cite{marion2023less} and semi-supervised learning algorithms \cite{killamsetty2021retrieve}. In these studies, the pruned subsets provided results with low or imperceptible loss of accuracy while requiring less data and computation time. In materials research, coreset selection has also been applied, for example, by using farthest point sampling to select training structures for crystal structure prediction \cite{wengert2021data} \REV{or by employing atomic structure featurization and clustering to sample minimal training data from molecular dynamics (MD) trajectories \cite{qi2024robust, sivaraman2020machine}.}

Active learning is another data reduction strategy. Generally, it refers to machine learning approaches that minimize training data volumes by optimal training data selection policies \cite{settles.tr09}. In computational materials science applications, active learning can be used to augment existing datasets \cite{zhang2023atomistic} or create new datasets from scratch \cite{gubaev2019accelerating, zhang2019active}. For example, Gaussian approximation potential (GAP) models trained on data generated by DFT or \textit{ab-initio} \REV{MD} have been improved with active learning \cite{sivaraman2020machine}. A similar process can be utilized for different kinds of models, such as committee neural networks \cite{schran2020committee}. Another application is Bayesian optimization structure search (BOSS), which uses active learning to construct potential energy surfaces with minimal computational effort \cite{todorovic2019bayesian}. BOSS has recently been applied to the study of perovskite materials' properties \cite{jinnouchi2019phase, li2023structural}.

\REV{Coreset selection and active learning have both been utilized individually in materials science applications, but their combination remains largely unexplored. In this work, our objective is to combine the two methods to create optimal datasets for training data-efficient ML models for structure relaxation.} To achieve this goal, we propose a three-step data generation scheme that minimizes the number of required DFT calculations. First, we generate a large pool of structures, and use clustering methods to select a diverse set of initial single point data. Then, we employ active learning to add structurally optimized data to the set, and finally use clustering again to prune the dataset. We test our approach by applying it to the aforementioned \ce{CsPb(Cl/Br)3} dataset and compare the results to our previous data generation method based on random sampling to assess performance. Additionally, we demonstrate the efficiency of our new data generation scheme for the inorganic ternary perovskite alloy \ce{CsSn(Cl/Br/I)3}. Ternary mixing of the X-site elements in a $2\times 2\times 2$ perovskite supercell provides significant configurational complexity that serves as an excellent test case for our data generation methodology. By generating the dataset, we aim to show that the scheme is widely applicable to different perovskite materials. An ML model fitted on the generated data would facilitate screening of the full ternary alloy space for stable materials candidates, but is the subject of future work.


In this article, we introduce our efficient data generation strategy and present its performance for reducing a preexisting dataset and for generating a new dataset. The rest of the paper is organized as follows. In Section \ref{methodology}, we go through the proposed approach step by step and establish the ML model used for predictions. Tests performed with the existing \ce{CsPb(Cl/Br)3} data are presented in Section \ref{binary results}. In Section \ref{sec:ternary_results}, we go through the process of applying the method to generate a novel \ce{CsSn(Cl/Br/I)3} dataset and present the results of ML predictions made on said data. In Section \ref{discussion}, we discuss these findings and outline future work, such as possible improvements and applications. Finally, in Section \ref{conclusions} we conclude with a summary.

\begin{figure*}
    \centering
    \includegraphics{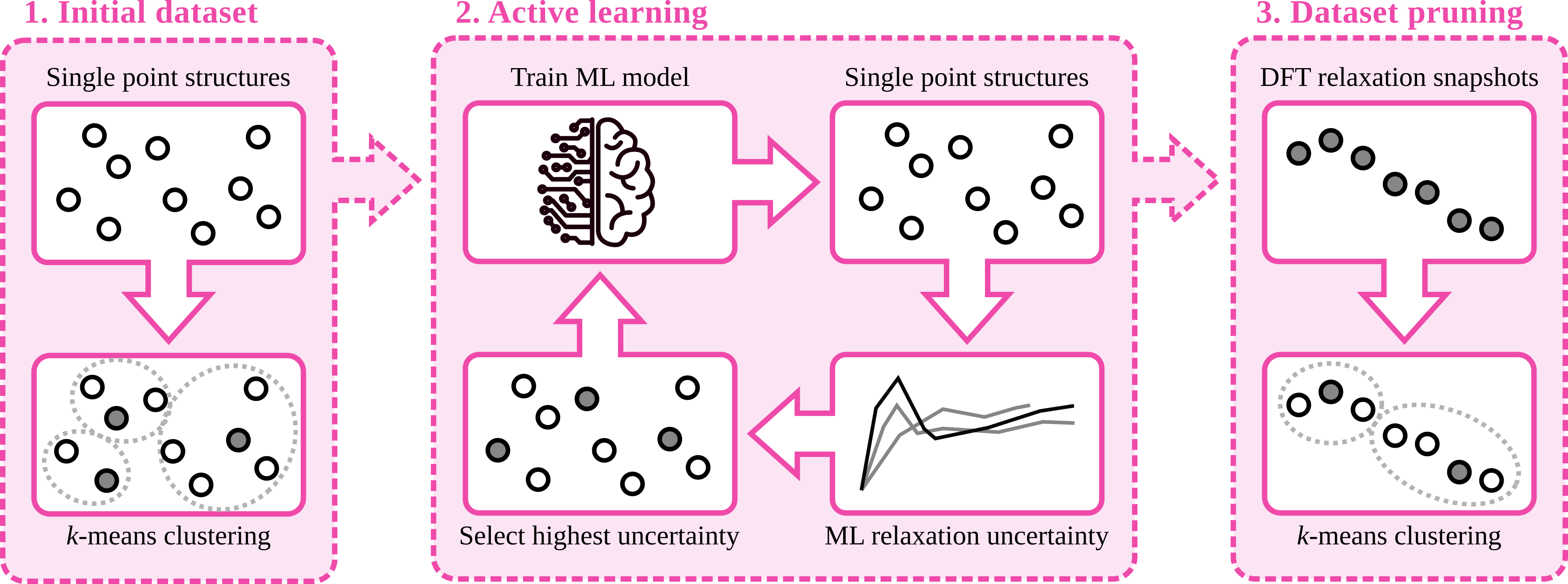}
    \caption{Workflow of the improved data generation schema in three parts.}
    \label{fig:recipe}
\end{figure*}

\section{Methodology}
\label{methodology}

In this section, we introduce our data generation process depicted in FIG. \ref{fig:recipe} step-by-step. Our approach has three parts: generating an initial dataset, using active learning to improve structural relaxation accuracy, and pruning. Each of these steps is explained in general terms in the following subsections. Computational details of data generation will vary depending on the application, and hence will be elaborated on in the respective sections of the two aforementioned datasets.

\subsection{Machine learning model}

The ML model we use in this article consists of a descriptor, which creates vector representations of the atomic structures, and a regression method that maps the vectorized structures into the corresponding energy values. As descriptor we use the many-body tensor representation (MBTR) \cite{huo2022unified} as implemented in DScribe \cite{Himanen/etal:2020, Laakso/etal:2023}, which encodes geometric features such as interatomic distances and angles as discretized Gaussian distributions. Based on earlier work \cite{stuke2019chemical, laakso2022compositional}, we conclude that the $k=2$ term for the inter-atomic distances already provides the desired accuracy while keeping the computational cost low.

To estimate the accuracy of our predictions during active learning, we also trained a Gaussian process regression (GPR) model. For the same kernel, GPR and KRR are equivalent, except GPRs predict distributions instead of scalars. When the KRR and GPR models are trained with the same data and same hyperparameters, the mean of the GPR distribution is identical to the KRR prediction, while the standard deviation ($\sigma$) quantifies the uncertainty of the prediction. We do not perform force evaluations with GPR, because its implementation is slower than our KRR model. Both our MBTR-KRR and MBTR-GPR models have been used in related work before \cite{laakso2022compositional, fang2024machine}.

\subsection{Initial dataset}
\label{1}

The first step in our workflow is to generate a dataset of single point structures to be used as initial training data for the ML models. This dataset should span the structural space of interest to prevent the need for extrapolation in areas with few training points. Additionally, the atomic structures chosen for the dataset should facilitate efficient learning for the ML model. Constructing such a dataset for structurally complex materials, such as perovskites, is a challenging task and the configurational complexity of perovskite alloys aggravates this further. Our solution is to first sample broadly, incorporating perovskite structures across all desired lattice types and varying degrees of octahedral tilting. We include a large number of different randomized alloy configurations sampling the composition space uniformly. To further enhance data diversity, we displace the atoms from their ideal lattice sites. In this way, we generate an arbitrarily large structure pool across the perovskite alloy space, from which we then select structures for DFT labeling with $k$-means clustering.

After a large number of atomic structures has been sampled, we generate MBTR vector representations for all structures. $k$-means clustering uses the Euclidean metric to quantify the similarity between two MBTR representations and thus between the corresponding atomic structures. To ensure similarly sized clusters from which it would then be straightforward to select the same number of structures every time, a minimum cluster size can be set using a constrained implementation of $k$-means clustering \cite{bradley2000constrained, Levy-Kramer_k-means-constrained_2018}. \REV{More analysis on the chosen clustering method, including a comparison of results with other clustering algorithms, is provided in Sec. S1 of the supplementary material (SM) \cite{supplement}.} After clustering, we randomly select an equal number of structures from each cluster for which we perform DFT calculations. These structure-energy pairs form the initial dataset of single point structures. 

\subsection{Active learning}
\label{2}

In the second step of our data generation workflow, the initial dataset is used as the starting point for active learning. While the goal in step 1 was to generate a maximally diverse dataset spanning the relevant structure space, the aim now is to generate specific additional data to improve ML based structure optimization.

We train MBTR-KRR and MBTR-GPR models on the initial data. A number of additional single point structures are generated uniformly across the composition space to be used as starting points for MBTR-KRR geometry relaxations using the BFGS algorithm \cite{fletcher2000practical}. For each structure along the relaxation trajectory, we compute the uncertainty with the MBTR-GPR model. Next, we pick certain structures according to the predicted uncertainties and perform DFT relaxations. The exact acquisition strategy depends on the application, and hence more details about our choices are provided in the following sections. The resulting trajectory data is added to the training dataset. Then, the ML model is re-trained with the updated dataset and the active learning loop is repeated until a desired ML relaxation accuracy is reached.

\subsection{Dataset pruning}
\label{3}

By their very nature, the relaxation trajectories include very similar structures. In the third and final step of our data generation scheme, we therefore reduce this redundancy by utilizing $k$-means clustering a second time, similarly to step 1. The relaxation trajectory data generated in step 2 is clustered and an equal number of points is picked from each cluster to form a smaller, representative dataset of relaxation snapshots, which is then combined with the initial single point data to form the final training dataset for the ML model. Since for all of the relaxation data points DFT energies have already been calculated, this pruning step simply reduces the size of the final dataset to decrease model execution times, but does not save DFT calculations.

\section{Validation for \NoCaseChange{\ce{CsPb(Cl/Br)3}}}
\label{binary results}

We tested our data generation method on a precomputed dataset of \ce{CsPb(Cl/Br)3} perovskite structures and their DFT total energies \cite{laakso2022compositional}. This dataset includes \num{10000} single point structures and 8014 relaxation snapshots. The single point data consists in equal parts of four different lattice types ($Pnma, I4/mcm, P4/mbm, Pm\bar{3}m$). The Cl/Br concentrations of the single point structures have been randomized but their distribution across the composition range is uniform. The relaxation data includes structure snapshots from 200 relaxation trajectories, 50 in each phase. 

In this section, we present the tests we performed for each step of the workflow. In Section \ref{bin_1}, we tested the initial dataset generation by comparing subsets selected by $k$-means clustering against a random selection. In Section \ref{bin_2}, we tested active learning by adding the relaxation trajectories with the most uncertain predictions into the training set of all single point structures, as opposed to adding random trajectories. And finally in Section \ref{bin_3}, we again compared $k$-means clustering with randomized pruning. We visualize the comparisons with learning curves demonstrating the effect of the modified data generation strategies on the learning process of the models.

\subsection{Initial dataset}
\label{bin_1}

\begin{figure}
    \centering
    \includegraphics{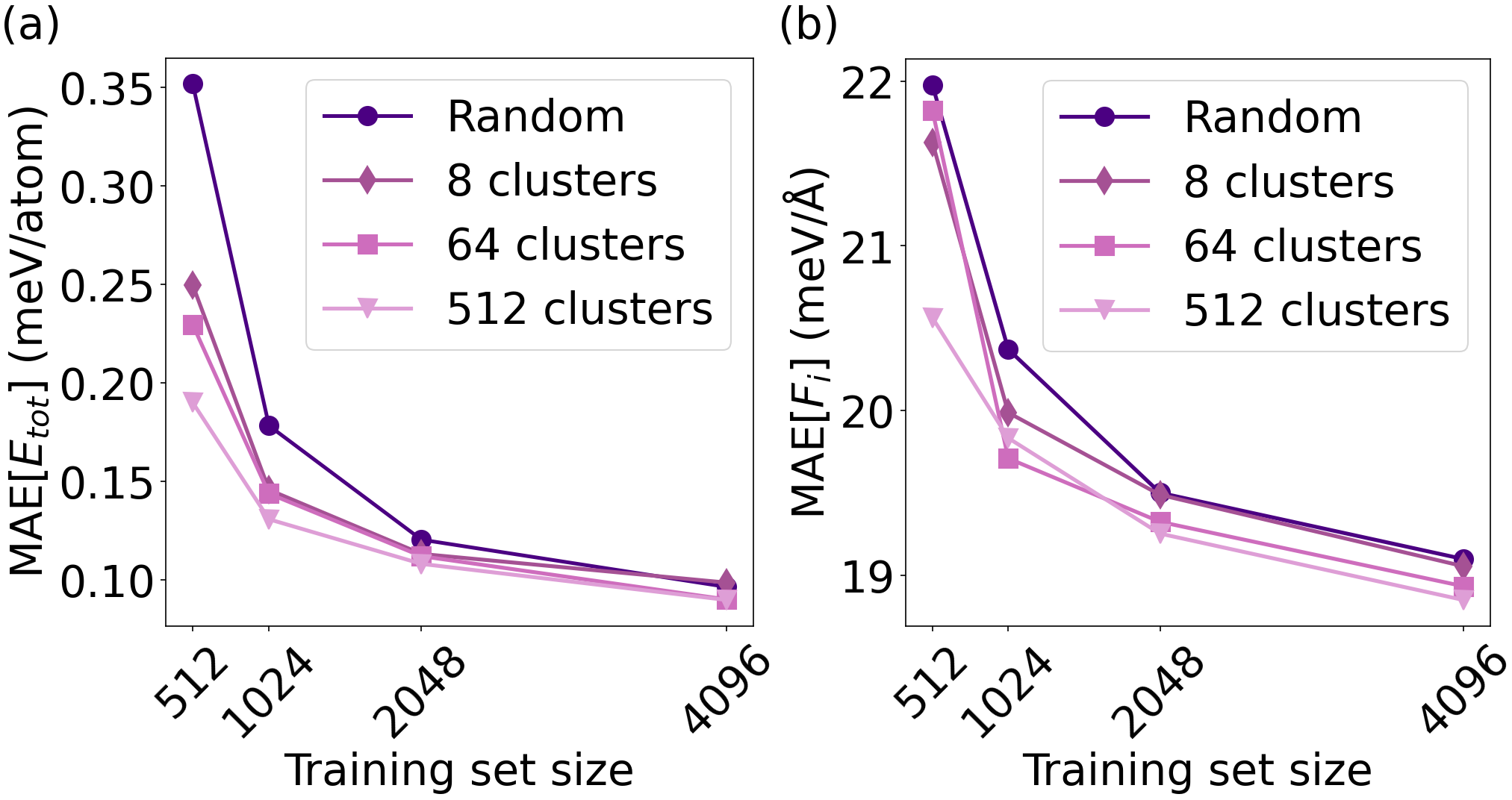}
    \caption{\ce{CsPb(Cl/Br)3} ML model prediction mean absolute errors during the initial dataset generation tests with increasing training set sizes and different numbers of clusters. Learning curves for single point energy (a) and force (b) predictions.}
    \label{fig:single}
\end{figure}


We assessed the effectiveness of $k$-means clustering for generating the initial dataset by applying it on 75\% of the \num{10000} available single point structures. Learning curves were plotted by keeping the clusters constant and selecting an increasing number of points from each cluster to form training sets of different sizes for the ML model. Mean absolute errors (MAEs) of energy and force prediction for the fitted models were then evaluated on the remaining 25\% of the single point data, with the final results being the mean of three randomized test-train splits. Moreover, we repeated the test with three different cluster counts ranging from 8 to 512, and compared the results to selecting training data randomly.

With $\sim$4000 structures used as training data, models using random selection and the clustering method both reach energy prediction errors of around 0.1 meV per atom, as seen in FIG. \ref{fig:single}. However, models trained with the cluster-selected structures reach these low errors much faster. In fact, the lowest value of the random model can be achieved with roughly 20\% less data using clustering. Force predictions exhibit a similar behaviour, with the lowest MAEs of 18.8 meV/Å and 19.1 meV/Å for clustering and random selection, respectively. Clustering saves 50\% data in this case. In general, a higher number of clusters tends to give better results for both energy and force predictions.

\begin{figure}
    \centering
    \includegraphics{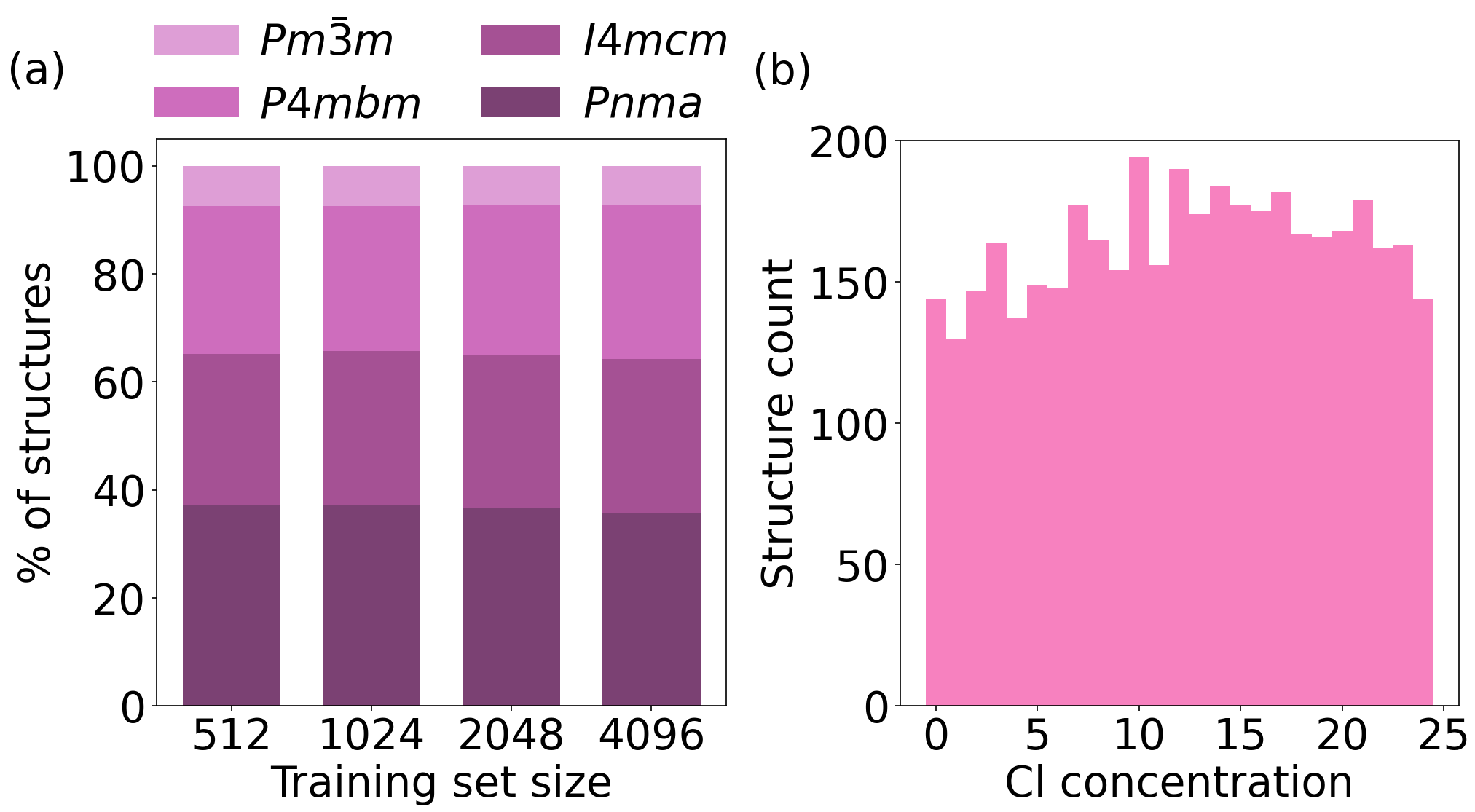}
    \caption{Details of selected \ce{CsPb(Cl/Br)3} structures for 512 clusters. (a) Phase distributions of single point structures for increasing training set sizes. (b) Cl concentrations for training set size 4096.}
    \label{fig:data}
\end{figure}

FIG.~\ref{fig:data} presents our analysis of the structures that were selected by the clustering method for 512 clusters. The distribution between the four phases stays consistent as the training set size increases, with slightly more $Pnma$ structures and significantly fewer $Pm\bar{3}m$ phases being selected. The Cl concentration, plotted here for dataset size of 4096, shows a convex shape with more structures being selected around the middle, despite the original dataset having uniform concentration distribution. More details for different cluster counts are presented in Sec. \REV{S2} of the SM \cite{supplement}.

\subsection{Active learning}
\label{bin_2}

\begin{figure*}
    \centering
    \includegraphics{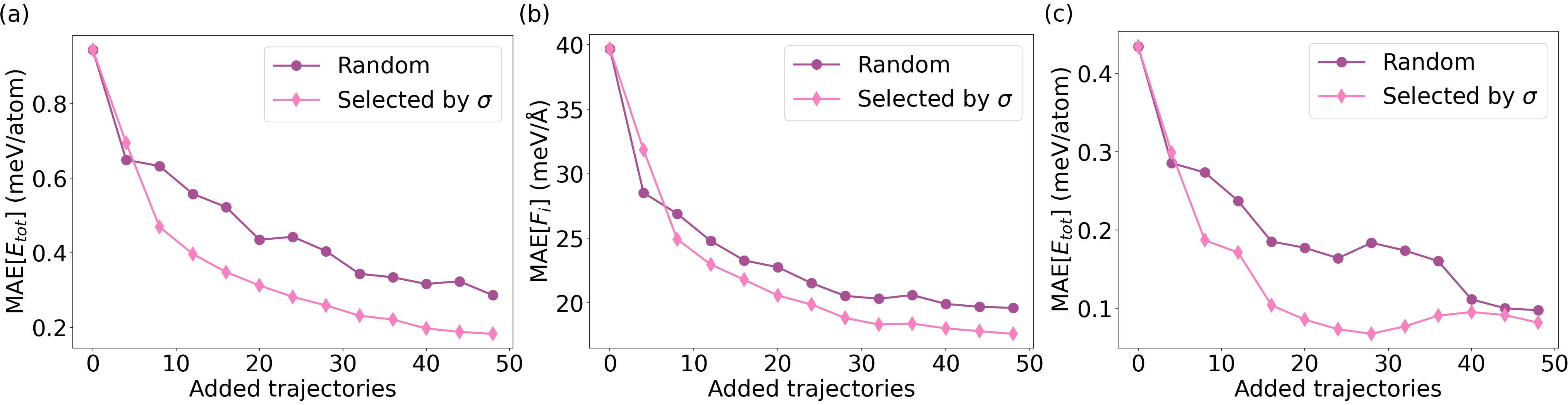}
    \caption{Active selection of \ce{CsPb(Cl/Br)3} relaxation trajectories: learning curves for (a) energy and (b) force predictions and (c) ML relaxation energies of the $Pnma$ phase.} 
    \label{fig:active}
\end{figure*}

We employed the relaxation data from the \ce{CsPb(Cl/Br)3} dataset to emulate the active learning step of the model training workflow. First, we trained an initial ML model using all \num{10000} single point structures. Subsequently, we divided the relaxation data into two equal parts: 100 relaxation trajectories available for training data augmentation and 100 for model testing. During each iteration of the active learning loop, we used the ML model to relax the 100 structures in the training set starting from the same initial geometries as used for the DFT relaxations. Then we selected the trajectory with the highest maximum uncertainty $\sigma$ along the entire ML relaxation trajectory in each of the four phases. The corresponding DFT relaxation snapshots of these four trajectories were added to the training data and the ML model was retrained. The loop was repeated 12 times to achieve a total of 48 added relaxation trajectories out of the 100 available for training. After every iteration, in order to monitor the model performance, we relaxed the 100 test structures with the ML model starting again from the initial DFT relaxation geometry, and compared the ML relaxed energies to the DFT relaxation results to obtain errors for structure relaxation. Additionally, error rates for energy and force predictions were estimated by using all structure snapshots from the 100 test DFT trajectories as testing data. We repeated the whole active learning process five times with different randomized train-test splits of the relaxation data. The final learning curves were computed as the mean of the five repetitions. For comparison, we repeated the process by selecting trajectories randomly, although also uniformly across the four phases.

The prediction errors converge much faster with the active learning method than with random sampling, as FIG.~\ref{fig:active} demonstrates. At 48 added trajectories, the MAE for energy predictions is 0.18 meV/atom for active learning and 0.29 meV/atom for random sampling. We obtain similar results for force predictions with 17.6 meV/Å for active learning and 19.6 meV/atom for random selection. Expressed in terms of data saving, active learning achieves the same accuracy as random sampling with half as much data for energy and force predictions. Although less consistent, improvements can also be observed for the relaxation predictions. Particularly for the $Pnma$ phase shown in FIG.~\ref{fig:active} the active learning model reaches lower errors much faster, resulting in more than 50\% data saving.

\subsection{Dataset pruning}
\label{bin_3}


Finally, we tested the third step of the data generation workflow, in which we reduce the redundancy of the DFT relaxation data through dataset pruning. We fitted the ML model with a training set consisting of all \num{10000} single point structures and relaxation snapshots selected via $k$-means clustering from 100 DFT trajectories. Energy and force prediction errors were again evaluated on a test set of structure snapshots from the remaining 100 DFT relaxation trajectories. Since the clustering tests in step 1 indicated that high cluster counts are optimal, we clustered the relaxation data into as many clusters as the intended number of included relaxation snapshots, selecting one structure from each cluster for the training set. To obtain learning curves, we varied the number of clusters from 200 to 2000. This repeated clustering of the data does not pose a computational problem, as even for a new dataset, we would already have DFT labels calculated for all data points. The final errors presented here are the mean of two-fold cross-validation, with the two halves of the relaxation data used once for training and once for testing. We also conducted pruning using random data selection and compared both approaches to a model trained without pruning, incorporating all the data from 100 DFT relaxation trajectories.

\begin{figure}
    \centering
    \includegraphics{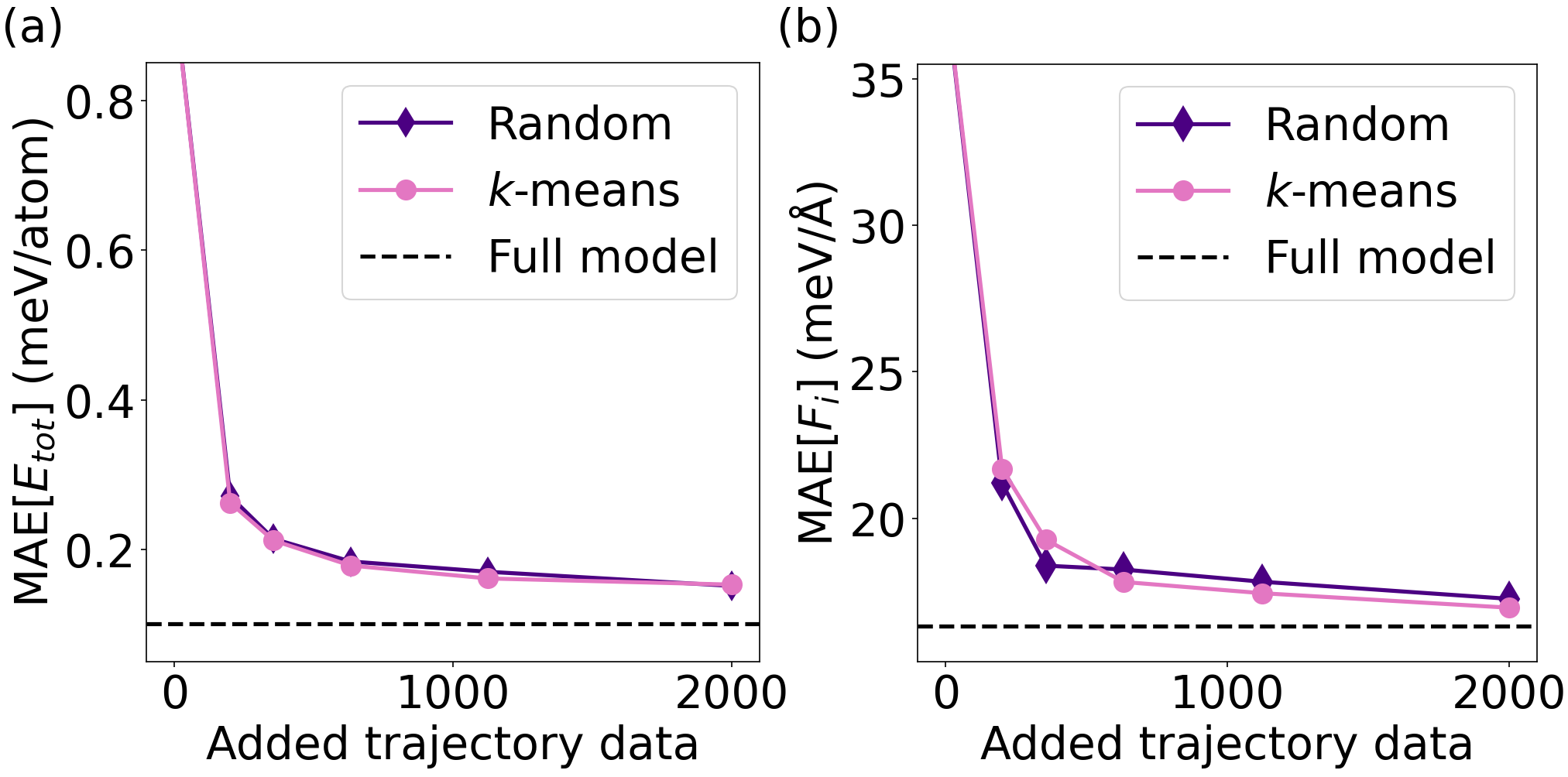}
    \caption{\ce{CsPb(Cl/Br)3} dataset pruning: the x-axis enumerates the relaxation snapshots added to the \num{10000} single point structures. Learning curves for (a) energy and (b) force predictions.}
    \label{fig:pruning}
\end{figure}

Using clustering in pruning the relaxation data gives small but consistent improvements over random sampling, as shown in FIG.~\ref{fig:pruning}. Both energy and force predictions start converging very quickly with increasing data. For example, after adding only $\sim$1 100 snapshots, MAEs for energy predictions drop to 0.170 meV/atom and 0.161 meV/atom for random selection and clustering, respectively. Similarly, force errors are 17.8 meV/Å for random selection and 17.4 meV/Å for clustering. For reference, the lowest possible error obtained with all structures of trajectories would be 0.1 meV/atom for the energies and 16.3 meV/Å for forces, and is marked by the dashed line in the figure.

\section{\NoCaseChange{\ce{CsSn(Cl/Br/I)3}} dataset generation}
\label{sec:ternary_results}

After testing our data generation approach with the preexisting \ce{CsPb(Cl/Br)3} binary perovskite alloy data, we applied the methodology to train a data-efficient ML model for the \ce{CsSn(Cl/Br/I)3} ternary alloy . We applied the main lessons that we learned from the earlier tests, but made minor modifications to the methodology due to the more complex materials space of the ternary alloy and the fact that we were not anymore limited by the static precalculated dataset.

DFT calculations for data generation were performed using the all-electron code FHI-aims \cite{blum2009ab, havu2009efficient, Xinguo/implem_full_author_list,levchenko2015hybrid}. We applied the same computational settings as in our previous \ce{CsPb(Cl/Br)3} study, employing the Perdew-Burke-Ernzerhof exchange-correlation functional for solids (PBEsol) \cite{perdew2008restoring}, the zero-order regular approximation for the relativistic effects (ZORA) \cite{lenthe1993relativistic}, standard FHI-aims tier-2 basis sets, ``tight'' grid settings, and a $\Gamma$-centered $4\times4\times4$ $k$-grid for Brillouin-zone integration. In support of open science, we made all relevant calculations available on the Novel Materials Laboratory (\mbox{NOMAD}) \cite{nomad} and Zenodo \cite{zenodo}. All the codes used for the \ce{CsSn(Cl/Br/I)3} dataset generation are available in a GitLab repository \cite{gitlab}.

\subsection{Initial dataset}

At the single point data generation stage, we followed the same steps as for the \ce{CsPb(Cl/Br)3} dataset. We generated 100 structures for all Cl/Br/I compositions and the same four space groups ($Pm\bar{3}m$, $P4/mbm$, $I4/mcm$ and $Pnma$) in $2\times 2\times 2$ \ce{CsSn(Cl/Br/I)3} supercells. The configuration of halide atoms in each generated structure was randomized. Atomic positions and lattice vectors were scaled according to Vegard's law with random deviations added to Cs and Sn positions, octahedral tilting angles, height/width ratio and volume of the cell, and the angle between the $\mathbf{a}$ and $\mathbf{b}$ lattice vectors. Generating 100 atomic structures for all four space groups and 325 halide compositions resulted in a data pool of \num{130000} atomic structures. Finally, we generated another two structures per composition and lattice type to obtain a dataset of 2600 atomic structures for model testing.

We then used clustering to select the atomic structures for DFT labeling. Following our \ce{CsPb(Cl/Br)3} model study, we opted for higher cluster counts. After computing the MBTR vectors for all \num{130000} structures, we therefore clustered all structures into 2000 clusters with a minimum of 20 structures in each cluster. We then performed single point DFT calculations in batches of 2000 structures selected randomly with one from each cluster. After each batch finished computing, we refitted our MBTR-KRR model to monitor the convergence of the energy and force predictions on the test set. The ML model hyperparameters were optimized with the Bayesian optimization code BOSS \cite{todorovic2019bayesian} following the methodology detailed in Refs.~\citenum{laakso2022compositional} and \citenum{Stuke/Rinke/Todorovic:2021}. The full list of optimized hyperparameter values can be found in Sec. \REV{S3} of the SM \cite{supplement}.

\begin{figure}
    \centering
    \includegraphics{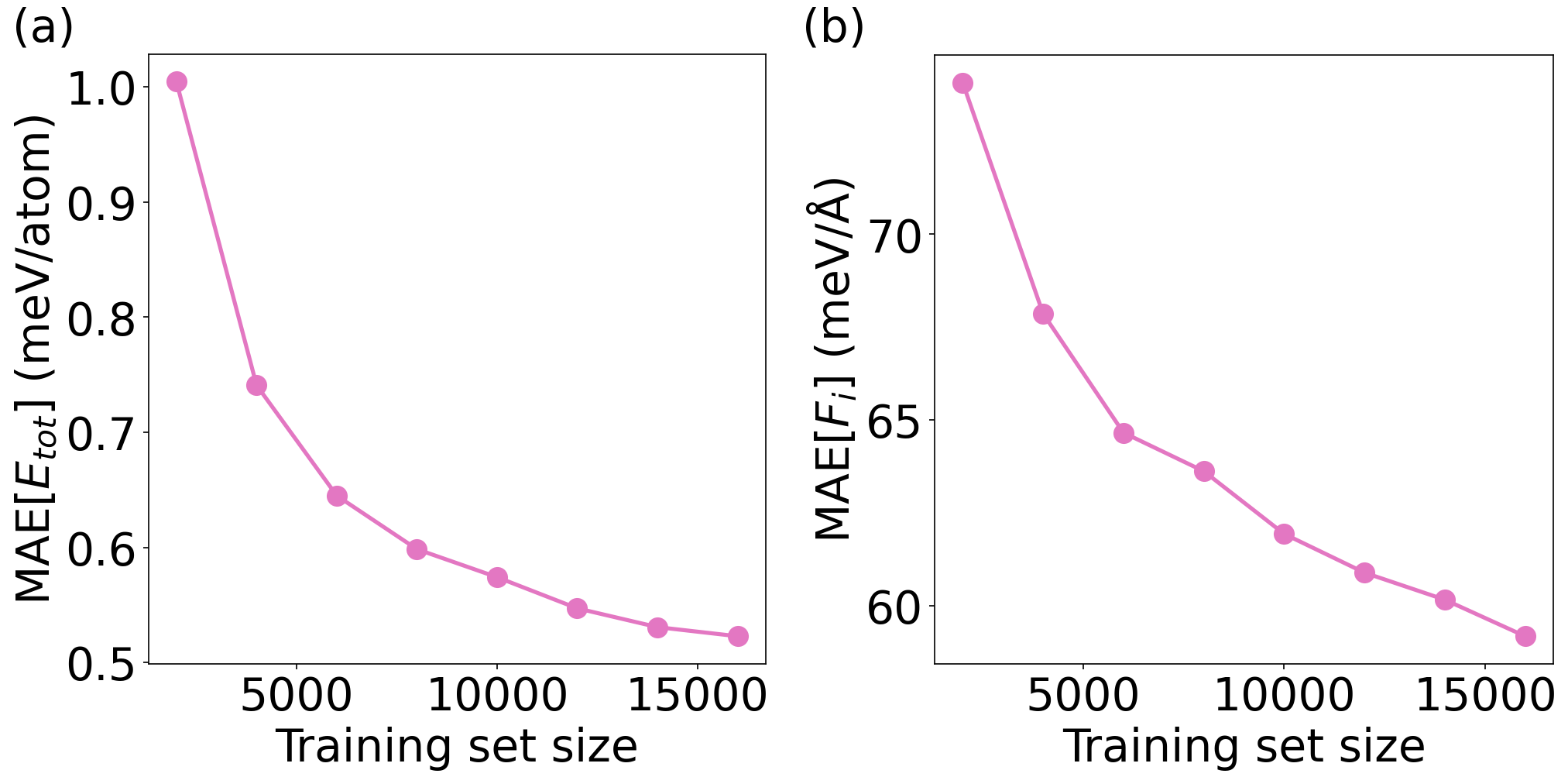}
    \caption{Single point \ce{CsSn(Cl/Br/I)3} data: learning curves for (a) single point energy and (b) force predictions.}
    \label{fig:ternary_sp}
\end{figure}

The resulting learning curves are shown in FIG.~\ref{fig:ternary_sp}. Both energy and force errors decrease rapidly with increasing training set size. After adding \num{16000} training structures the energy MAE has converged to approximately \SI{0.5}{meV/atom}. The force MAE has reached \SI{59}{meV/\AA} and is still decreasing with added batches. We nonetheless decided to stop single point DFT data generation at this point due to the diminishing returns for energy predictions. 

\subsection{Active learning}

\begin{figure}
    \centering
    \includegraphics{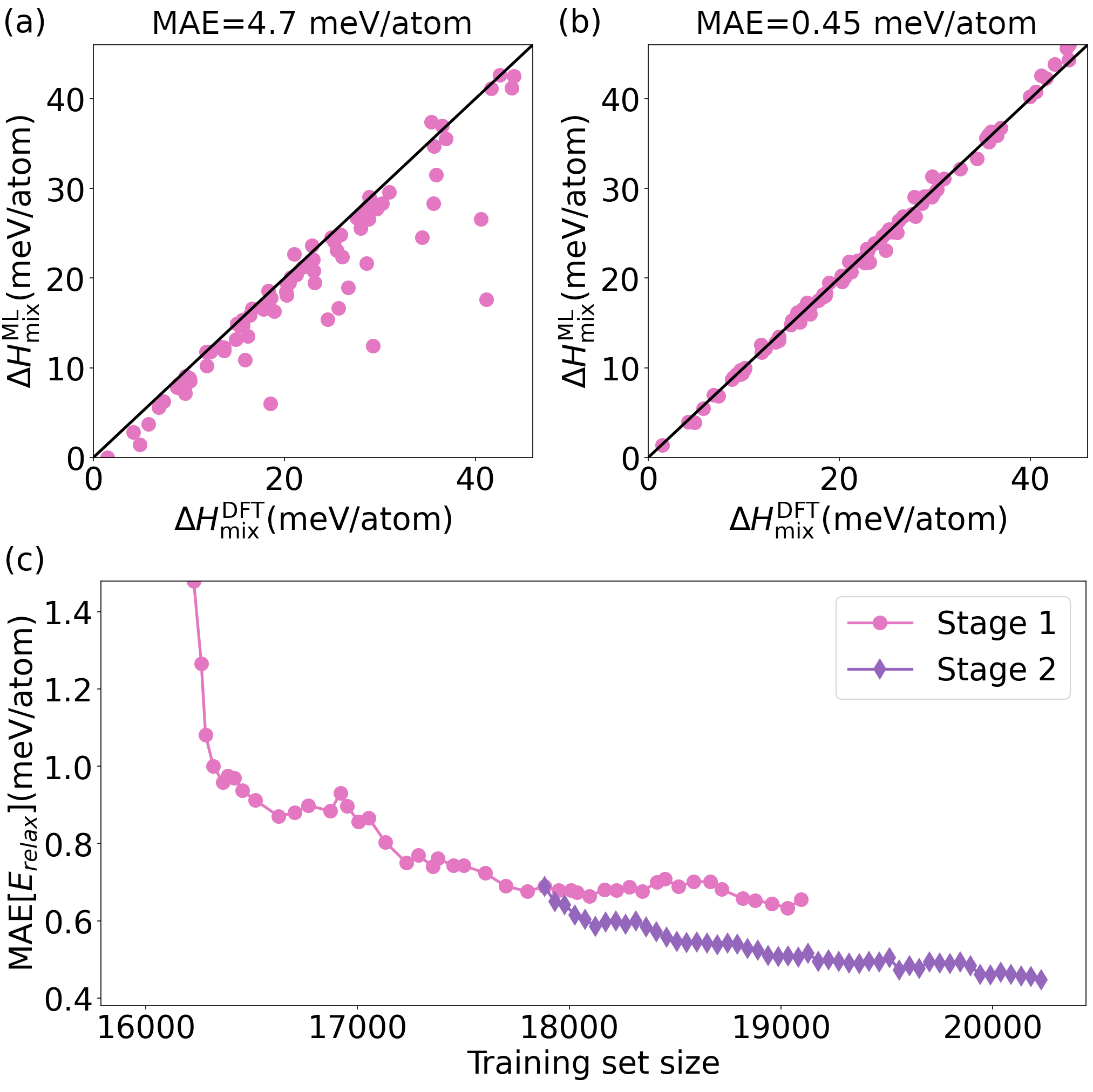}
    \caption{Active learning for \ce{CsSn(Cl/Br/I)3}: (a) Enthalpies of mixing ($\Delta H_{\text{mix}}$) for \ce{CsSn(Cl/Br/I)3} structures relaxed with DFT and the initial \num{16000} data point ML model. Shown are only the 88 relaxations that converged with the ML model. (b) Same comparison with the final ML model. (c) Energy MAE of ML structure relaxations during the active learning.}
    \label{fig:ternary_al}
\end{figure}

After having generated the initial dataset of \num{16000} atomic structures, we tested the corresponding MBTR-KRR model by relaxing 100 \ce{CsSn(Cl/Br/I)3} structures. The initial structures for the relaxations were generated in a similar way to the single point structures of the initial dataset but without introducing the random deviations to the atomic positions and cell shape. All four phases were equally represented in this test set, with 25 initial structures each. To prevent phase transformations during relaxation, the halide positions and lattice parameters were constrained to the symmetries dictated by the respective space groups of the four phases, while the Cs and Sn positions were allowed to vary freely. For comparison, we relaxed the same structures with DFT and compared the resulting energies with the ML predictions. The results of this test are shown in FIG.~\ref{fig:ternary_al}a. Only 88\% of the ML relaxations converged within 200 relaxation steps and the MAE of the ones that did was \SI{4.7}{meV/atom}.

To improve the ML relaxations, we devised a two-stage active learning protocol that targets i) relaxation convergence and ii) accuracy around the equilibrium structures. At the first stage, we generated 25 structures per phase in each iteration of the active learning loop, in the same way as was done for the relaxation test described in the previous paragraph. We then relaxed all generated structures with the ML model and monitored the prediction uncertainty with the corresponding GPR model. The two ML relaxation trajectories that exceeded an uncertainty threshold of \SI{0.5}{meV/atom} with the least relaxation steps were selected from each phase, and the relaxation was continued with DFT from the step that exceeded the limit. We set the force convergence limit for the DFT relaxations to a relatively loose value of \SI{0.1}{eV/\AA} to keep the relaxations short. Finally, all the structure snapshots from the DFT relaxation trajectories were added to the training set and the KRR and GPR models were refitted for the next iteration of the active learning loop. With increasing data, the ML model improves and eventually fewer than two relaxations per phase exceeded the uncertainty limit. When this happened, we decreased the uncertainty limit for the corresponding phase by 20\% before proceeding with the next iteration of the active learning loop.

At the second stage of the active learning protocol, we modified the acquisition strategy to improve the accuracy of the ML model near equilibrium structures. The ML relaxations were now run until the forces were smaller than \SI{0.005}{eV/\AA}. We then computed the uncertainties for the corresponding structures with the GPR model and selected the two structures per phase with the highest uncertainties. DFT relaxations were launched from these structures and run for exactly five steps. All structures obtained in this way were added to the training data.

We monitored the performance of the ML model by repeating the relaxation test described above in each iteration of the active learning loop. The resulting learning curves are shown in FIG.~\ref{fig:ternary_al}c. Initially, the MAE drops rapidly with active learning iterations. Progress then slows and starts to level out after approximately 2000 structures have been added to the training data. At this point, we changed to the second stage of the active learning protocol. The MAE drops more rapidly again and continues to decrease with increasing data. The scatter plot comparing DFT and ML relaxed energies of individual structures is shown in \ref{fig:ternary_al}b. With the final ML model, all the test relaxations converge and the MAE is only \SI{0.45}{meV/atom}.

\section{Discussion}
\label{discussion}

For the fixed sized \ce{CsPb(Cl/Br)3} dataset, we observed that clustering with a larger number of clusters produced better results. This is to be expected, because we chose the training structures randomly from each cluster. For a fixed dataset size, a large number of smaller clusters increases diversity, whereas a smaller number of larger clusters approaches random sampling. However, there are some downsides and limitations to using a large number of clusters. The computational complexity of $k$-means clustering increases with increasing cluster count. Moreover, having fewer clusters simplifies monitoring the ML model training process. The number of clusters sets the resolution of the learning curve, with the smallest data unit being the total number of clusters. This precision allows for earlier termination of the training process once the desired accuracy is achieved exactly. \REV{Based on our findings, we recommend using 500 clusters for binary alloys and \num{2000} clusters for ternary alloys. However, the optimal cluster count will ultimately depend on the choice of ML model and the target accuracy of the fitted model.}

As for the structures that get selected by clustering, we observed that structures from the lower symmetry phases are picked more often than others. This is sensible and helpful, as these structures have more structural variety and are thus harder for the ML model to predict. For the Cl concentrations that get picked, the convex shape seen in FIG.~\ref{fig:data}b is also intuitive, since the concentrations in the middle have many more possible atomic configurations. However, this shape only emerges with small cluster sizes, with larger ones actually leading to the opposite effect (see FIG. \REV{S2} of the SM \cite{supplement}). This change in clustering behaviour may be one of the reasons why larger cluster counts yielded better results in our tests.

The validation results of the active learning step presented in Section \ref{bin_2} are also promising. However, it should be noted that this test on preexisting data does not entirely reflect a practical application of generating a new dataset from scratch, since eventually both active learning and random selection converge to the same values as we run out of possible new structures to pick. Nonetheless, we later on successfully utilized our data generation method to create a new dataset, with minor adjustments made to the active learning loop in order to fit the new type of data.

In the final dataset pruning step clustering only improves the results by very little compared to random selection. Regardless of the pruning method, it is clear that the full relaxation trajectories are not needed. Pruning at this step should be considered on a case-by-case basis, since all of the DFT calculations have already been completed and the only possible gain is the reduction of the ML fitting time and the size of the final ML model. Depending on the model, there may be good reasons to decrease the training set size by limiting the number of redundant structures, such as faster prediction times.

The model for the ternary \ce{CsSn(Cl/Br/I)3} perovskite  that was trained with the initial dataset of \num{16000} structures predicted energies accurately but exhibited a relatively high force prediction error of \SI{59}{meV/\AA}. This is not surprising as we only used energies for training the model. The force prediction error could likely be decreased with ML models that use both energies and forces for training. The high force prediction led to poor relaxation accuracy for the initial model, but the active learning protocol reduced this error to a remarkably low value of \SI{0.5}{meV/atom}, which is comparable to the error that we had previously achieved for the structurally less complex binary perovskite. The improvement in relaxation accuracy was achieved with little added data (about 20\% more than the single point structures in the initial data set). Considering the effectiveness of our active learning approach, the composition of the final dataset suggests that we could have stopped the initial dataset generation earlier. The possibility of accelerating learning with ML models leveraging atomic forces for training, together with the prioritization of active learning over initial data generation, presents a pathway toward even greater data efficiency in future applications of our workflow.

In this work, we used a well established MBTR-KRR ML model. Our choice was based on the good experience made in our previous work \cite{laakso2022compositional,fang2024machine}, but in principle our data generation scheme can be used for any ML method that produces energies and forces and provides access to uncertainties. However, changing the ML model might require adjustments to the data generation workflow. For instance, using the MBTR descriptor to calculate structural distances for both clustering and the ML model kernel may create synergies that are absent if different distance metrics are used in the two steps. \REV{The optimal number of clusters for the initial dataset generation may also change -- fewer clusters should be used with models that have higher learning rates to avoid exceeding the necessary number of data points.} Additionally, the active learning step is highly dependent on the chosen ML algorithm, especially the computational speed of fitting the model. In this work, we added only a small number of trajectories in each active learning loop, but for a more complex model with slower fitting times an approach with fewer loops and more added data per loop should be adopted instead. Computational cost considerations also matter for the final pruning step: for models whose prediction time increases with increased training data, the pruning step is much more important.

\REV{We designed the data generation scheme in this work specifically to train ML models for structure optimization, but it could also be adapted for training MD potentials. When generating data from scratch for ML MD potentials, a common strategy has been to sample uncorrelated structures from \textit{ab-initio} MD simulation trajectories. Our initial data generation method that employs clustering provides a promising alternative, particularly for alloy systems, where the need to capture a wide range of alloy configurations in the initial dataset renders MD trajectory sampling inefficient. Another fundamental weakness of the MD sampling method is that, due to the requirement of selecting only uncorrelated structures, most of the expensive DFT calculations performed during the MD simulations are essentially wasted from a model training perspective. Our approach circumvents this problem by not incorporating any MD trajectory data during the initial dataset generation.}

\REV{The main modification required for adapting our scheme to MD potential training is in the active learning step, which would need to incorporate MD simulations instead of structure relaxations. The fundamental principles would remain the same: the ML model runs an MD simulation until the prediction uncertainty exceeds a threshold, at which point the energy of the structure is computed with DFT and added to the training set. This approach is already widely used and has been shown to be effective. Another necessary change to the workflow would be in model validation. In this work, we assessed ML model accuracy by comparing relaxation energies obtained with ML and DFT, but for MD potentials, validation should involve large-scale MD simulations to ensure the reliability of the model.}

In this study, we developed a machine learning model to expedite the relaxation of atomic structures in the \ce{CsSn(Cl/Br/I)3} alloy. This model enhances the efficiency of scanning alloy compositions to identify stable material candidates. However, due to the high configurational complexity of the ternary alloy, thoroughly exploring the alloy space remains time-consuming, even with the ML model. Additionally, Sn-based perovskite alloys, compared to their Pb-based counterparts, generally have a higher enthalpy of mixing, which increases the relative contribution of entropy to alloy stability and makes it challenging to gain sufficient insight from internal energy analysis alone. We are currently working on methods to address these computational challenges by incorporating advanced configuration sampling techniques and ML-based approaches for determining configurational and vibrational entropy, and we will present these advancements in future publications.

\section{Conclusions}
\label{conclusions}

In this work, we developed an efficient data-generation scheme to facilitate machine learning model training for structural relaxations of perovskite alloys. We tested our scheme on an existing \ce{CsPb(Cl/Br)3} perovskite dataset, showing that our data pruning and active learning methods can reduce the required training data by 20\% during initial dataset generation and by approximately half during relaxations, without compromising prediction accuracy for energies, forces, or geometries. We then applied this scheme to generate a new dataset for the more complex \ce{CsSn(Cl/Br/I)3} ternary perovskite alloy. Using active learning to generate relaxation snapshots proved highly effective, resulting in an ML model with remarkably low prediction error for structure relaxations. Our results highlight the potential of strategic dataset generation to enhance ML model training efficiency, paving the way for computational studies of perovskite alloys and other complex materials with quantum mechanical precision.

\section*{Acknowledgements}

We thank Pascal Henkel, Marcelo Marques, Jingrui Li, and Milica \mbox{Todorovi\'{c}} for fruitful discussions. This work was supported by the Academy of Finland through Projects No. 334532 and 352861 and COST actions CA18234 and CA22154. For computational resources, we further wish to acknowledge CSC-IT Center for Science, Finland, and the Aalto Science-IT project.


\bibliography{apssamp}

\end{document}


\title{Supplementary Material: \\
Efficient dataset generation for machine learning perovskite alloys}

\author{Henrietta Homm}
\affiliation{%
 Department of Applied Physics, Aalto University, P.O. Box 11100, 00076 Aalto, Finland
}%

\author{Jarno Laakso}
\affiliation{%
 Department of Applied Physics, Aalto University, P.O. Box 11100, 00076 Aalto, Finland
}%

\author{Patrick Rinke}
\affiliation{%
 Department of Applied Physics, Aalto University, P.O. Box 11100, 00076 Aalto, Finland
}%
\affiliation{Physics Department, Technical University of Munich, Garching, Germany}
\affiliation{Atomistic Modelling Center, Munich Data Science Institute, Technical University of Munich, Garching, Germany}
\affiliation{Munich Center for Machine Learning (MCML)}

\date{\today}

\maketitle
\tableofcontents

\clearpage

\section{Comparison of clustering methods}
\label{s0}

\begin{figure}
    \centering
    \includegraphics[width=0.5\textwidth]{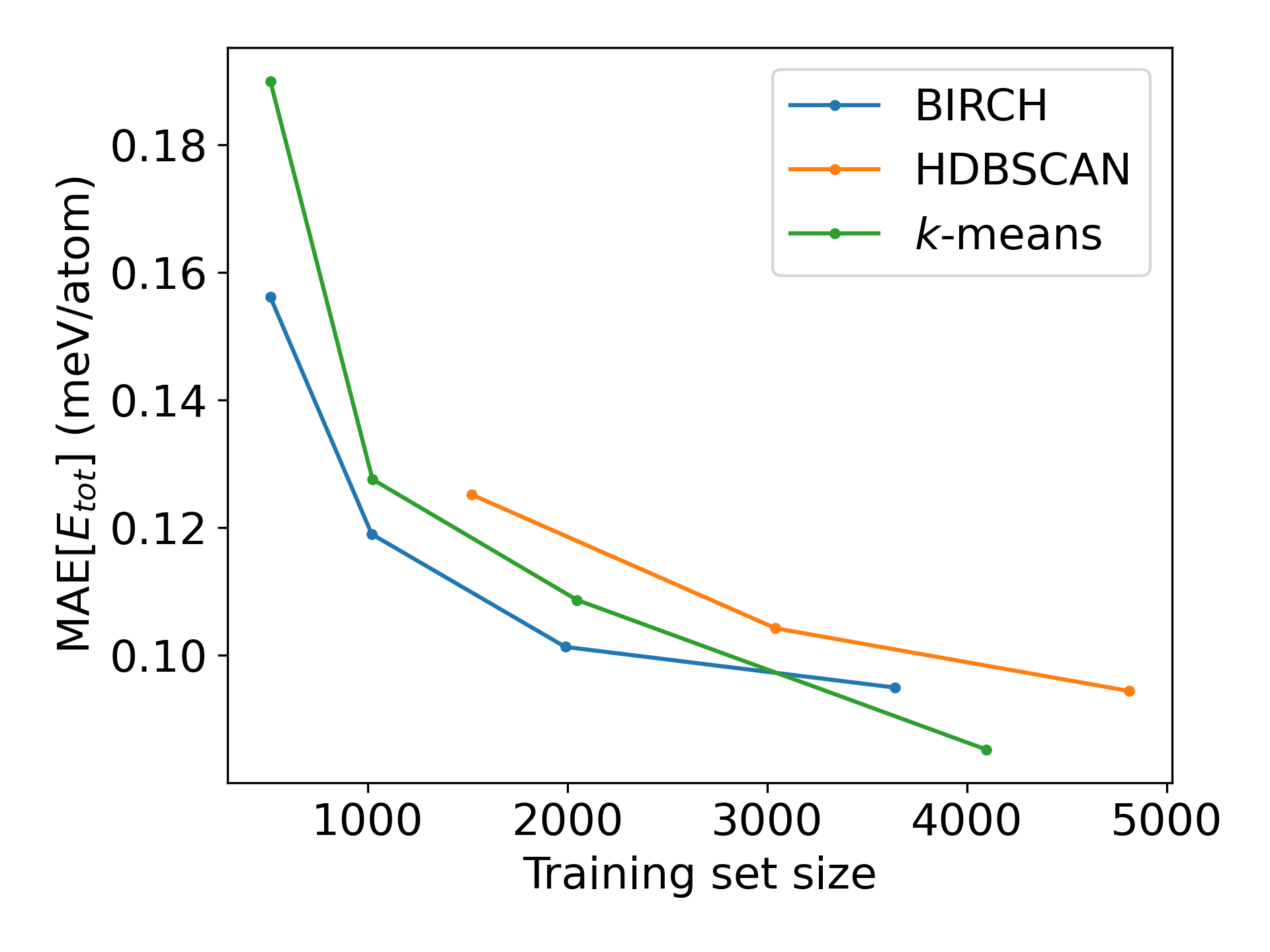}
    \caption{Mean absolute errors of \ce{CsPb(Cl/Br)3} ML model predictions: single point energies during initial dataset generation tests with different clustering algorithms and an increasing amount of data selected with each method.}
    \label{fig:methods}
\end{figure}

We opted to use constrained $k$-means as the clustering method in our workflow. A few others were also tested, and the results of the best-performing alternatives, BIRCH (Balanced Iterative Reducing and Clustering using Hierarchies) \cite{zhang1996birch} and HDBSCAN (Hierarchical Density-Based Spatial Clustering of Applications with Noise) \cite{campello2013density} are presented in FIG. \ref{fig:methods}. These tests were performed, similarly to the validation results presented in the article, with the set of \num{7500} single point \ce{CsPb(Cl/Br)3} structures used as training data and the remaining \num{2500} as a test set. With these results, it remains that the relatively simple $k$-means algorithm is highly suitable for the clustering task, and we can identify three main reasons for why this is the case. For one, in the constrained implementation of $k$-means, both the number of clusters and the minimum cluster size can be set as parameters, which allows for precise control. Secondly, it is easily scalable to large amounts of data and many clusters, and finally, unlike DBSCAN and its derivative HDBSCAN, it does not have outlier removal. Removing many outliers may turn out to be harmful to the resulting model predictions, as the unique outlier structures contain important information and removing them leads to extrapolation.

\section{Initial dataset selection}
\label{s1}

\begin{figure}
    \centering
    \includegraphics[width=\textwidth]{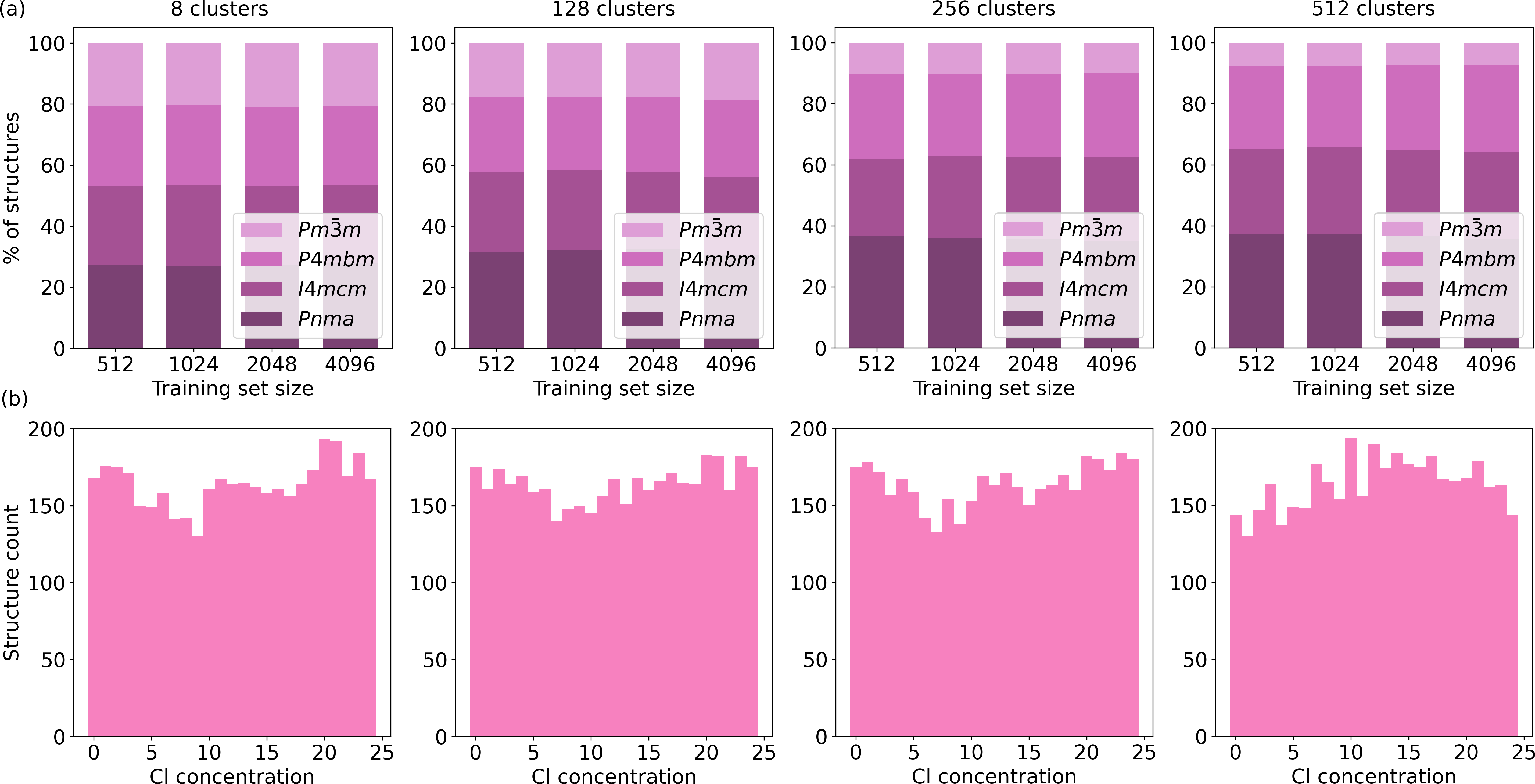}
    \caption{Details of \ce{CsPb(Cl/Br)3} structures that get selected using different cluster counts. (a) Phase distributions of single point structures in increasing training set sizes. (b) Cl concentrations in training set size 4096.}
    \label{fig1}
\end{figure}

In the article, we provided some discussion on which single-point structures get picked by our clustering method. FIG. \ref{fig1}a presents phase distributions with different cluster counts for all training set sizes. As the clusters get smaller, the phase distribution shifts to prefer the more complex phases on all dataset sizes. With smaller clusters, there's less randomness in the structures that get picked from them. As the simpler phases end up in large clusters with little variation within them compared to smaller clusters of more complex structures far away from each other, it then follows that more of those end up in the final selection.

FIG. \ref{fig1}b shows the Cl/Br concentration distributions of the structures picked by the clustering algorithm with different cluster counts for the full set of \num{4096} structures. These distributions also exhibit a significant change when increasing the number of clusters, as the selection method goes from preferring the ends of the concentration range to preferring the middle. The convex shape in the smaller cluster numbers is rather counterintuitive, as the concentrations in the middle have many more possible atomic configurations. Since there's more structural variation in the middle, it's easy to think that more of those should be selected. One possible explanation for this is that there's fewer and larger clusters near the center of the concentration space, fewer clusters resulting in less data being picked compared to the smaller clusters more separate from each other at the ends. Since a smaller amount of clusters means larger ones everywhere, it would then make sense that this effect turns around when the clusters get smaller across the entire feature space.

\begin{figure}
    \centering
    \includegraphics[height=0.45\textheight]{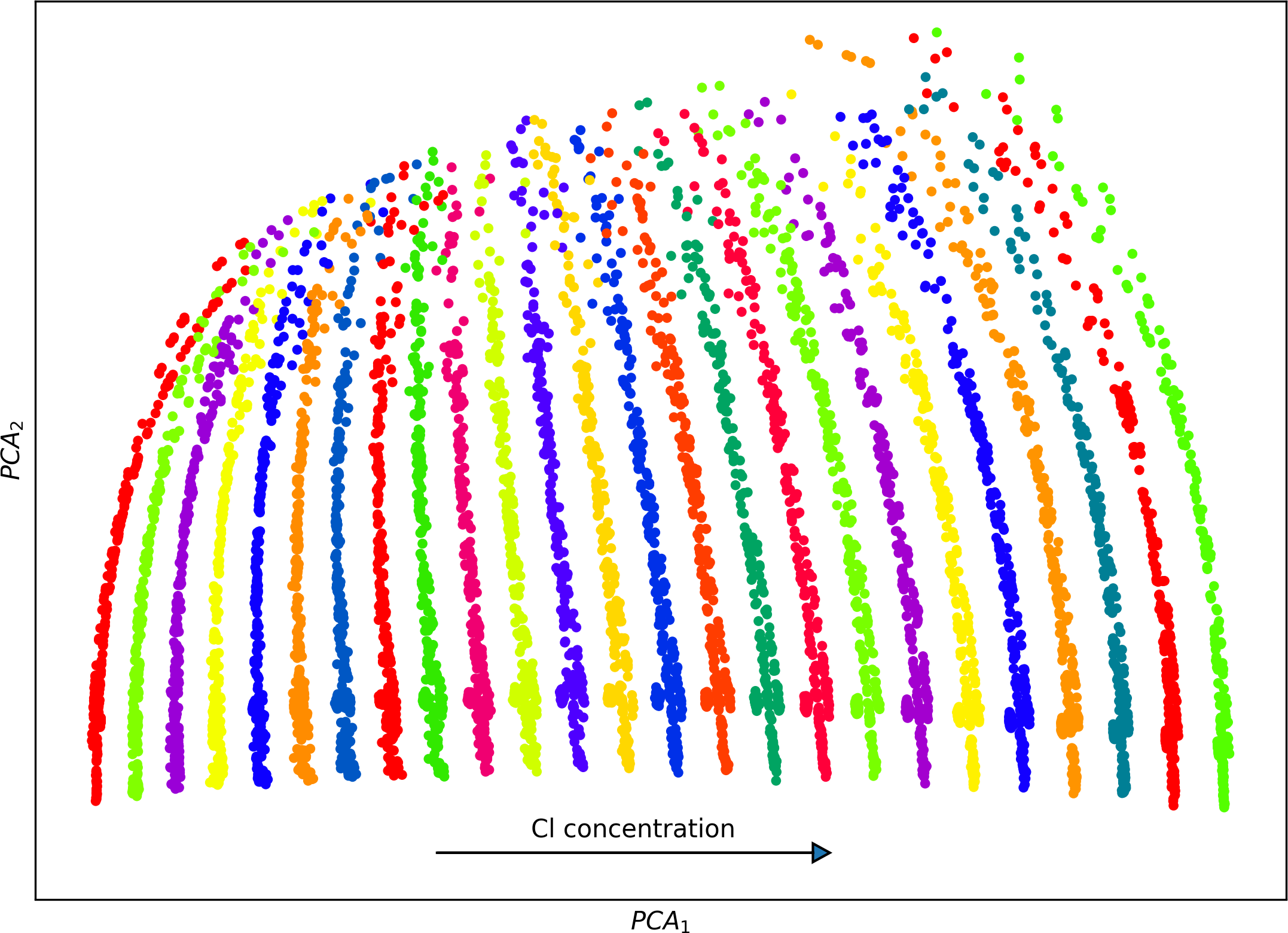}
    \caption{Cl/Br concentrations in single point training data as they show up in data reduced into two dimensions using PCA.}
    \label{fig2}
\end{figure}

\begin{figure}
    \centering
    \includegraphics[height=0.45\textheight]{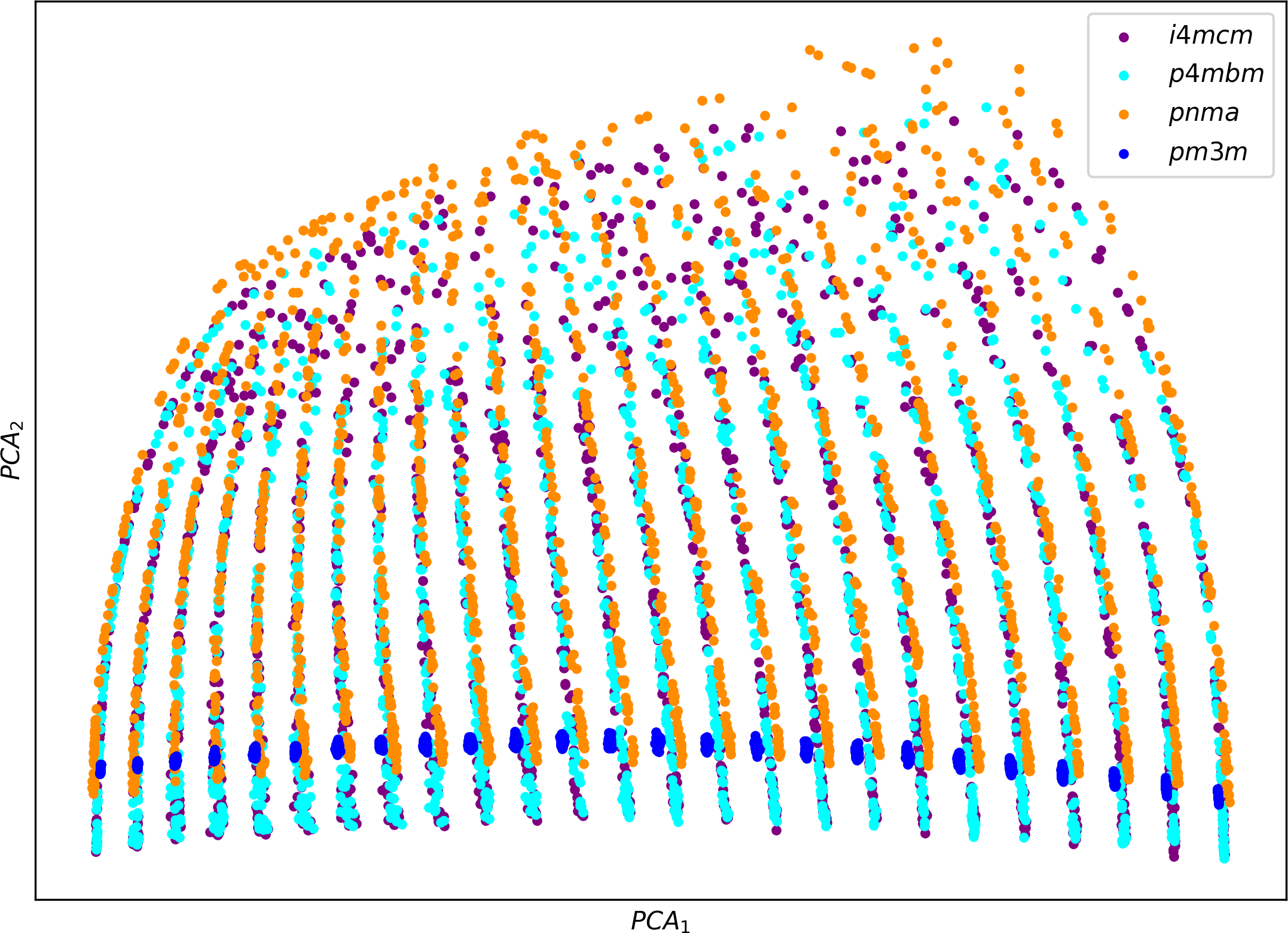}
    \caption{The four phases in single point training set, flattened by dimensional reduction.}
    \label{fig3}
\end{figure}

\begin{figure}
    \centering
    \includegraphics[height=0.45\textheight]{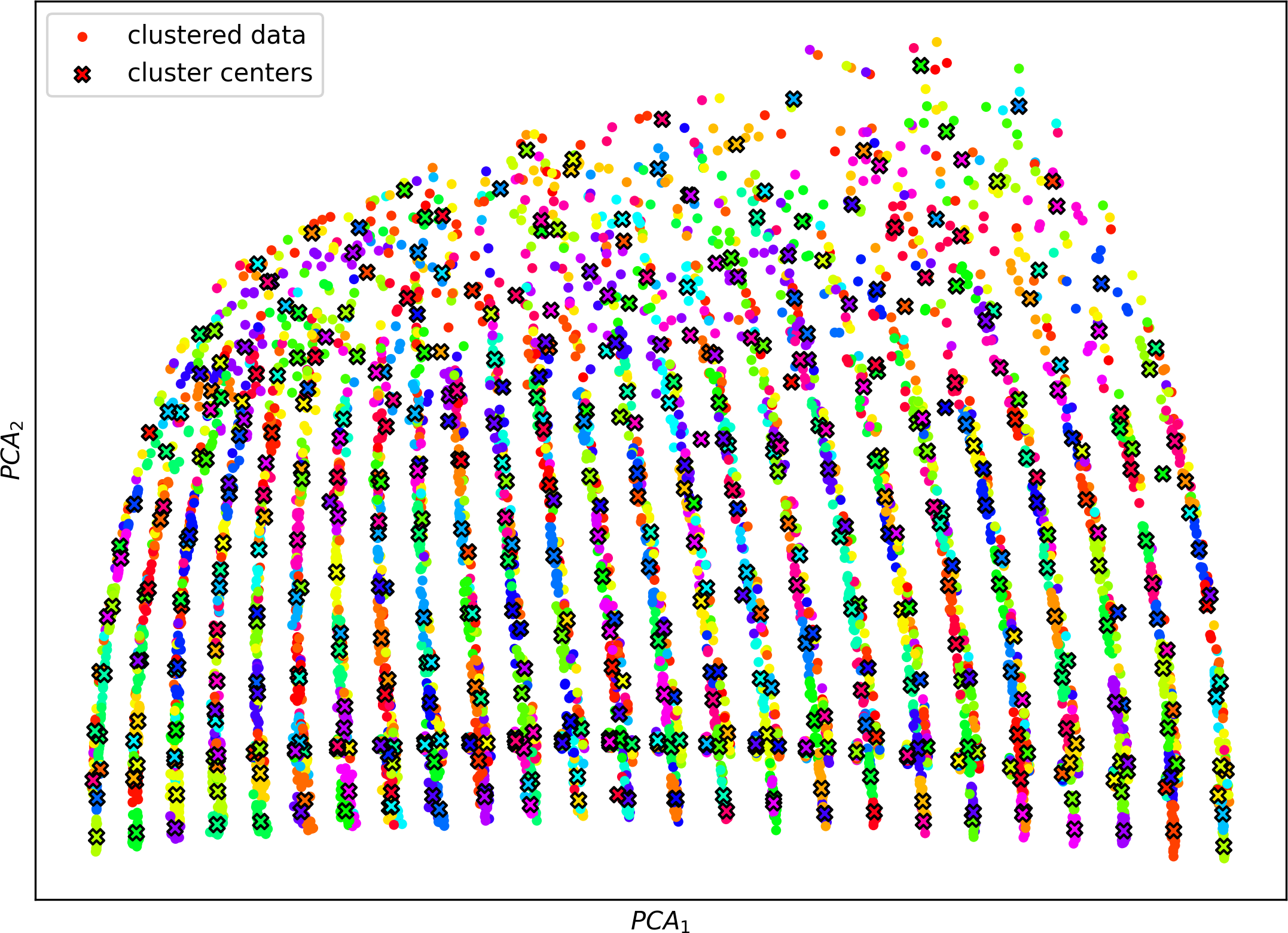}
    \caption{Clusters and cluster centers of a single run with 512 clusters.}
    \label{fig4}
\end{figure}

For a closer look into how the clusters are formed, we used principal component analysis (PCA) for dimensionality reduction on the 500-dimensional MBTR vectors representing atomic structures. This way, we can plot the training dataset of 4096 single-point structures in two dimensions. The first PCA-component relates strongly to the Cl concentration of the structure, as seen in FIG. \ref{fig2}. The phases plotted in FIG. \ref{fig3} suggest that the second component relates to structural complexity with regards to atomic displacement of Cs, Cl and Br, as structures with phase $Pm\Bar{3}m$ have none of this octahedral tilting and can be found in a nearly straight line. Finally, FIG. \ref{fig4} presents the results of a single clustering instant with 512 clusters. From this we can see the reason for the phase distributions seen in FIG. \ref{fig1}a. The $Pm\Bar{3}m$ phase structures very close together in the feature space get clustered into very few clusters, resulting in less data points being picked from them.

When viewing this analysis, it is important to keep in mind that the clustering was not executed on the dimensionally reduced PCA output, but rather the original MBTR vectors describing the structures in full. Here the clustering results are only flattened afterwards for visualization.

\clearpage

\section{Machine learning model hyperparameters}

The machine learning (ML) model used in this work has eight hyperparameters. $x_{\text{min}}$ and $x_{\text{max}}$ determine the extents for the discretization grid for MBTR functions, while $N$ is the number of grid points. $\sigma$ is the standard deviation of the Gaussian distributions in MBTR. $r_{\text{cut}}$ and $w_{\text{cut}}$ are the MBTR weighting parameters, where $r_{\text{cut}}$ determines the cutoff radius and $w_{\text{cut}}$ the magnitude of weighting at the cutoff distance. $\alpha$ is the regularization parameter of the KRR model, and $\gamma$ determines the length scale of the Gaussian kernel in KRR. Detailed information on the ML model and all of its hyperparameters can be found in our earlier work \cite{laakso2022compositional}.

Hyperparameters $x_{\text{min}},$ $x_{\text{max}}$ can be chosen based on the characteristics of the data, while $N,$ $r_{\text{cut}},$ $w_{\text{cut}},$ and $\alpha$ are trade-off parameters that were chosen so that the ML model can produce accurate but efficient predictions. The remaining two hyperparameters, $\sigma$ and $\alpha,$ were optimized with Gaussian optimization utilizing the method detailed in \cite{laakso2022compositional}.

The optimized hyperparameter values for the two ML models are shown in TABLE \ref{tab:hyperparameters}. Notably, the optimal $\sigma$ value is somewhat higher for the ternary perovskite model. The large difference in $\gamma$ is explained by the fact that in the ternary model, MBTR vectors were normalized with the number of atoms in the atomic structures. This rescaling of the feature space causes the optimal value for $\gamma$ to be $40^2$ times larger than it would be without the normalization of MBTR vectors.

\begin{table}[t]
    \centering
    \begin{tabular}{r|c|c}
  
         & \ce{CsPb(Cl/Br)3} & \ce{CsSn(Cl/Br/I)3} \\
         \hline
         $x_{\text{min}}$ (\si{\AA^{-1}}) & -0.1 & -0.1 \\
         $x_{\text{max}}$ (\si{\AA^{-1}}) & 0.6 &  0.6\\
         $N$ & 50 & 50\\
         $\sigma$ (\si{\AA^{-1}}) & \SI{1.752e-2}{} & \SI{4.279e-2}{} \\
         $r_{\text{cut}}$ (\AA) & 6.27 & 7.04 \\
         $w_{\text{cut}}$ & \SI{1.0e-3}{} & \SI{1.0e-3}{} \\
         $\alpha$ & \SI{1.0e-5}{} & \SI{1.0e-5}{}\\
         $\gamma$ & \SI{1.280e-4}{} & \SI{4.279e-2}{}\\
         \hline
    \end{tabular}
    \caption{Hyperparameters of the ML models for \ce{CsPb(Cl/Br)3} and \ce{CsSn(Cl/Br/I)3}.}
    \label{tab:hyperparameters}
\end{table}

\bibliography{supplement}